\documentclass[entropy,accept,oneauthor,12pt,a4paper]{mdpi}
\lastpage{\pageref*{LastPage}}
\doinum{10.3390/e1608xxxx}
\pubvolume{16}
\pubyear{2014}
\history{Received: 5 June 2014 ; in revised form: 4 August 2014 / Accepted: 4 August 2014 / \\Published: xx August 2014}

\usepackage{soul}

\Title{Is Gravity Entropic Force?}

\Author{Rongjia Yang}

\address[1]{%
College of Physical Science and Technology,
Hebei University, Baoding 071002, China; \\
E-Mail: yangrongjia@tsinghua.org.cn}

\corres{}

\abstract{If we assume that the source of thermodynamic system, $\rho$ and $p$, are also the source of gravity, then either thermal quantities, such as entropy, temperature, and chemical potential, can induce gravitational effects, or gravity can induce thermal effects. We find that gravity can be seen as entropic force only for systems with constant temperature and zero chemical potential. The case for Newtonian approximation is discussed.}


\begin{document}


\section{Introduction}

It has been recognized that there are profound connections between gravity and \mbox{thermodynamics \cite{Cocke,Bekenstein,Hawking,Davies,Unruh}}. Since then, these connections has been steadily becoming stronger. It has been shown that the entropy $S$ can be taken as the Noether charge associated with the diffeomorphism invariance of the theory \cite{Wald,Iyer}. In \cite{Jacobson}, the Einstein equation has been derived from the first law of thermodynamics. This attempt also has been investigated in modified gravity theories \cite{Eling,Elizalde,Brustein} and has been revisited in \cite{Makela}. In a wide class of spacetime, the field equations in both general relativity and Lovelock theories can be expressed as a thermodynamic identity near the horizon \cite{Padmanabhan,Paranjape,Kothawala} (see a \mbox{review \cite{Padmanabhan2010}}). In \cite{Gao2011}, the generalized Tolman--Oppenheimer--Volkoff equation is derived by using the maximum entropy principle to a charged perfect fluid, impling that there are fundamental relationships between general relativity and ordinary thermodynamics. The equations of motion for modified gravity theories, such as $F(R)$ gravity, the scalar-Gauss--Bonnet gravity, $F(\mathcal{G})$ gravity, and the non-local gravity, are equivalent to the Clausius relation in thermodynamics \cite{Bamba}. In \cite{Bracken}, the Einstein--Hilbert action can be constructed by minimizing the free energy. It was argued that the variation of the surface term evaluated on any null surface which acts a local Rindler horizon can be given a direct thermodynamic interpretation \cite{Parattu}. Gravity was explained as an entropic force caused by changes in the information associated with the positions of material bodies \cite{Verlinde}. Possible modifications and extensions to this interesting idea were \mbox{proposed \cite{Gao2010,Li,Cai,Hendi}}. All these studies were based on some assumptions, such as Unruh temperature, horizon, null surfaces, and so on. In \cite{Yang}, The thermal entropy density has been obtained for any arbitrary spacetime without assuming a temperature or a horizon, implying that gravity possesses thermal effects, or, thermal entropy density possesses effects of gravity.

Here we generalize the results in \cite{Yang} to the case of nonzero chemical potential. The results we obtained indicate that the changes of temperature, entropy, particles, and chemical potential will result in gravitational force, or gravitational force will induce changes of temperature, entropy, particles, and chemical potential.

\section{Relations between Gravity and Thermodynamics}

Both the energy density $\rho$ and the pressure $p$ play important roles in general relativity or thermodynamics. $\rho$ and $p$ are components of the stress-energy tensor in general relativity and are fundamental variables in thermodynamics. Let us begin with the first law of thermodynamics for fluids consisting of particles in curved spacetime
\begin{eqnarray}
\label{1st}
dE=TdS-pdV+\mu dN.
\end{eqnarray}
where $E$, $S$, and $N$ represent the total energy, entropy and particle number within the volume $V$, $\mu$, $T$, and $p$ are the chemical potential, the temperature, and the pressure of the perfect fluid, respectively. $dV=\sqrt{h}d^3x$ with $\sqrt{h}$ the determinant of the spatial metric. We take $c=G=1$ and use metric signature $(-,+,+,+)$ throughout this paper. Rewriting (\ref{1st}) in terms of densities
\begin{eqnarray}
\label{1ast}
d(\rho V)=Td(sV)-pdV+\mu d(nV),
\end{eqnarray}
we can easily get
\begin{eqnarray}
\label{1bst}
\rho dV+ Vd\rho=TVds+TsdV-pdV+n\mu dV+V\mu dn,
\end{eqnarray}
where $s$ is the entropy density and $n$ the particle number density. Applying Equation  (\ref{1st}) to a unit volume, we have
\begin{eqnarray}
\label{1cst}
d\rho=Tds+\mu dn.
\end{eqnarray}
Combining Equations (\ref{1bst}) and (\ref{1cst}), we obtain the integrated form of Gibbs--Duhem relation \cite{Gao2001}
\begin{eqnarray}
\label{1dst}
Ts=\rho+P-\mu n.
\end{eqnarray}
For $\mu=0$, Equation (\ref{1dst}) reduces to the thermal entropy density obtained in \cite{Yang}. Now, considering the Einstein equation in \cite{Wald1984}
\begin{eqnarray}
\label{E}
R_{\mu\nu}-\frac{1}{2}g_{\mu\nu}R+\Lambda g_{\mu\nu}=8\pi T_{\mu\nu},
\end{eqnarray}
and the stress energy tensor of the perfect fluid
\begin{eqnarray}
\label{T}
T_{\mu\nu}=g_{\mu\nu}p+(\rho+p)u_{\mu}u_{\nu},
\end{eqnarray}
we obtain
\begin{eqnarray}
\label{E1}
R-4\Lambda =- 8\pi (3p-\rho).
\end{eqnarray}
Using the $3+1$ decomposition of Einstein equation, we derive \cite{Wald1984,Gourgoulhon}
\begin{eqnarray}
n^{\mu}n^{\nu}R_{\mu\nu}+\frac{1}{2}R-\Lambda =8\pi \mathcal{E},
\end{eqnarray}
where $n^{\mu}$ is the unit normal vector field to the 3 dimension hypersurfaces $\Sigma$ and $\mathcal{E}=\Gamma^2(\rho+p)-p$ with $\Gamma$ the Lorentz factor. According to the scalar Gauss relation, one can get
\begin{eqnarray}
\label{3E}
\mathcal{R}+K^2-K_{ij}K^{ij}-2\Lambda =16\pi \mathcal{E},
\end{eqnarray}
where $\mathcal{R}$ is the Ricci scalar of the 3 dimension hypersurfaces $\Sigma$, $K_{ij}$ the extrinsic curvature tensor of~$\Sigma$, and $K$ the trace of the $K_{ij}$.
Combining Equations (\ref{E1}) and (\ref{3E}), we obtain the expression of $\rho+p$ in general relativity \cite{Yang}
\begin{eqnarray}
\label{ted1}
\rho+p=\frac{1}{4\pi(4\Gamma^2-1)} \left[\mathcal{R}+K^2-K_{ij}K^{ij}-\frac{1}{2}R \right].
\end{eqnarray}
The four dimension Ricci scalar, $R$, can be decomposed as \cite{Gourgoulhon}
\begin{eqnarray}
R=\mathcal{R}+K^2+K_{ij}K^{ij}-\frac{2}{N}\mathcal{L}_m K-\frac{2}{N}D_iD^{i}N,
\end{eqnarray}
where $\mathcal{L}_m$ is the Lie derivative along $\mathbf{m}$ of any vector tangent to $\Sigma$ and $D_i$ is the Levi--Civita connection associated with the metric of the 3 dimension hypersurfaces $\Sigma$. Then Equation (\ref{ted1}) can be expressed with three dimension spacial geometrical quantities as \cite{Yang}
\begin{eqnarray}
\label{ted3}
\rho+p &=& \frac{1}{8\pi(4\Gamma^2-1)} \left[ \mathcal{R}+K^2-3K_{ij}K^{ij}+\frac{2}{N}\mathcal{L}_m K+\frac{2}{N}D_iD^{i}N \right].
\end{eqnarray}
In order to relate thermodynamics with gravity, we must suggest a hypothesis. We first review Newton's equivalence principle so as to have a better understanding. When a particle falls freely in a gravitational field, the gravity is also the inertial force. This fact leads to Newton's equivalence principle that the inertial and the gravitational mass of a particle are equal. Like the case of Newton's equivalence principle, for any perfect fluid in spacetime the energy density $\rho$ and the pressure $p$ are both the sources of gravity and thermodynamic system. So we put forward a hypothesis that the source of thermodynamic system, $\rho$ and $p$, are also the source of gravity, namely
\begin{eqnarray}
\label{hyp}
(\rho+p)_{\rm gravitational~source}=(\rho+p)_{\rm thermal~source}.
\end{eqnarray}
For radiation, this hypothesis holds \cite{Yang}. In \cite{Jacobson}, the $\delta Q$ of thermodynamic system was assumed as the energy flux and then the Einstein equation was obtained. Analogous assumptions have also been suggested in \cite{Yang,Gao2011,Wald2013,Gao2013}, though these assumptions have not been aware of by the authors. With this hypothesis (\ref{hyp}) and combining Equations (\ref{1dst}) and (\ref{ted3}), we obtain
\begin{eqnarray}
\label{key}
Ts+\mu n &=& \frac{1}{8\pi(4\Gamma^2-1)} \times \left[ \mathcal{R}+K^2-3K_{ij}K^{ij}+\frac{2}{N}\mathcal{L}_m K+\frac{2}{N}D_iD^{i}N \right].
\end{eqnarray}
The terms of the left-hand side of this equation are thermodynamical quantities, while the terms of the right-hand side of this equation are geometrical quantities of the spacetime. Equation (\ref{key}) implies that gravity can induce thermal effects, or, thermal quantities, such as entropy, temperature, and chemical potential, can induce gravitational effects. To understand the physical significance of Equation (\ref{key}) well, we consider the Newtonian approximation: $g_{\mu\nu}=\eta_{\mu\nu}+h_{\mu\nu}$, with $h_{\mu\nu}\ll 1$. For matter with $p\simeq0$ (the pressure of a body becomes important when its constituent particles are traveling at speeds close to that of light, which we can exclude from the Newtonian limit by hypothesis), such as dust or dark matter, we have the Poisson equation $\nabla^2\varphi=4\pi \rho$ with $\varphi=-h_{00}/2$. Taking into account Gibbs--Duhem \mbox{relation (\ref{1dst})} and the hypothesis (\ref{hyp}), we obtain
\begin{eqnarray}
\label{key1}
\nabla^2\varphi=4\pi(Ts+\mu n).
\end{eqnarray}
Equation (\ref{key1}) (or (\ref{key})) relates gravity with thermodynamics. Variations in temperature, entropy, chemical potential, and particle number will lead to a variation in the potential $\varphi$ , and vice versa. Equation (\ref{key1}) (or (\ref{key})) also indicates that gravity is related to the entropy but is not entropic force. Only for systems with constant temperature and zero chemical potential, gravitational force is entropic force. These results confirm the arguments in \cite{Kobakhidze} that experiments with ultra-cold neutrons in the gravitational field of Earth disprove the speculations on the entropic origin of gravitation. In \cite{GaoS}, it also argued that the argument for the entropic origin of gravity is problematic.

\section{Conclusions}

We have shown that if we conjecture that the source of thermodynamic system, $\rho$ and $p$, are also the source of gravity: $(\rho+p)_{\rm gravitational~source}=(\rho+p)_{\rm thermal~source}$, thermal quantities, such as entropy, temperature, and chemical potential, can induce gravitational effects, or gravity can induce thermal effects. For Newtonian approximation, the gravitational potential is related to the temperature, entropy, chemical potential, and particle number, which implies that gravity is entropic force only for systems with constant temperature and zero chemical potential. For general case, gravity is not an entropic force. Whether the results obtained here can be generalized to the case of modified gravity, such as $F(R)$ \mbox{gravity \cite{Elizalde}} and $F(\mathcal{G})$ \mbox{gravity \cite{Bamba}}, is worthy of investigation. All the analyses have been carried out without assuming a specific expression of temperature or horizon. For a static system at thermal equilibrium in general relativity, the temperature of the perfect fluid may take the form, $T\sqrt{-g_{00}}=const.$, which is called the Tolman \mbox{temperature \cite{Tolman35,Tolman36,Rovelli}}. Whether the temperature in Equation (\ref{1st}) can be taken as Tolman temperature is also worthy of further investigation. The results we obtained confirm that there is a profound connection between gravity and thermodynamics.

%
\acknowledgements{Acknowledgments}
This study is supported in part by National Natural Science Foundation of China (Grant Nos. 11147028 and 11273010) and Hebei Provincial Natural Science Foundation of China (Grant No. A2014201068).


\newpage
\conflictofinterests{Conflicts of Interest}

The author declares no conflict of interest.

\bibliographystyle{mdpi}
\makeatletter
\renewcommand\@biblabel[1]{#1. }
\makeatother


\end{document}